\documentstyle[twocolumn,prl,aps,psfig]{revtex}
\tightenlines
\begin{document}
\draft
\title{Role of spin-orbit coupling in the far infrared absorption 
of lateral semiconductor dots}
\author{Manuel Val\'{\i}n-Rodr\'{\i}guez, Antonio Puente, and
Lloren\c{c} Serra}
\address{Departament de F\'{\i}sica, Universitat de les Illes Balears,
E-07071 Palma de Mallorca, Spain}
\date{December 10, 2001}
\maketitle
\begin{abstract}
We investigate the relevance of the spin-orbit coupling to the 
far-infrared 
absorption of two-dimensional semiconductor dots. Varying the strength 
of the Dresselhaus term, a mechanism feasible in experiment by changing 
the dot width, distinctive splittings of the Kohn peak as well as
additional low energy modes are predicted in a non-interacting model. 
Each mode has a spatial  
distribution of charge, perpendicular and in-plane spin densities that
correlate in a peculiar way 
with the frequency and polarization of the external field. 
We study the robustness of these features against electron-electron
interactions as well as the appearance of interaction-induced 
additional characteristics. 
\end{abstract}



The role of the electron spin-orbit (SO) coupling in the 
general properties of lateral semiconductor dots has attracted much 
interest in recent years. To a large extent this is motivated by the 
technological impact attributed to the so-called {\em spintronic} devices, 
whose operation principle exploits precisely the SO 
mechanism to control the spin transport \cite{Dat90}. 
Actually, spintronics is an incipient field whose fundamental
principles need a deeper understanding \cite{Wol01}.  
In this general context, experimental and 
theoretical investigations 
have very recently shown that the SO coupling  
affects the quantum transport and, more specifically, the  
conductance fluctuations of chaotic quantum dots in a
parallel magnetic field \cite{Fol01,Hal01,Ale01}.

The different SO contributions in semiconductor dots 
have been reviewed by Voskoboynikov {\em et al.} \cite{Vos01}.
In general, SO coupling has two contributions:
the Rashba term due to the nanostructure confinement 
potential; and the Dresselhaus term originating from the 
crystalline inversion asymmetry. In lateral dots 
the strength of the latter 
depends on the vertical width, which can be experimentally
varied with vertical electric fields. For an embedding
GaAs medium both Rashba and Dresselhaus parameters are 
reasonably determined \cite{Kna96}. 

In this Letter we study the influence of SO couplings on the dipole 
absorption of GaAs dots with a parabolic confinement. 
While the Rashba term provides a negligible 
effect we show that the Dresselhaus contribution leads to characteristic 
splittings of the Kohn modes and to the appearance of quasi-spin flip modes
in the far-infrared absorption (FIR). For each mode we find a correlation between
the spatial patterns of oscillating electron density, horizontal (in-plane) 
spin density, and vertical spin density. 
As in most of the literature treating 
the SO coupling \cite{Hal01,Ale01,Vos01},
we shall first consider 
a non-interacting model, using both analytical and numerical treatments.
Nevertheless, the role of the interaction will be also considered within 
self-consistent mean-field for Hartree and exchange-correlation 
contributions,
using a spinorial formalism of local-spin-density theory (LSDA)
\cite{Hei99}.

Within the effective Hamiltonian formalism and for a
quasi two-dimensional system  
the Rashba and Dresselhaus [for the standard (001) plane of GaAs]
contributions read, respectively, 
\begin{eqnarray}
\label{eq1}
{\cal H}_R &=& \frac{\lambda_R}{\hbar} \sum_{i=1}^{N}{  
\left[\, {\vec\sigma}\cdot\left(\,\nabla{V}({\bf r})\times{\bf P}\right)
\,\right]_i
} 
\nonumber\\
{\cal H}_D &=& \frac{\lambda_D}{\hbar} \sum_{i=1}^{N}{ 
\left[\, P_x\sigma_x-P_y\sigma_y\,\right]_i} \; ,
\end{eqnarray}  
where the $\lambda$'s control the coupling strengths, the $\sigma$'s refer to 
the Pauli matrices and $V({\bf r})$ stands for the total electrostatic potential.
In Eq.\ (\ref{eq1}), ${\bf P}$ gives the kinetic momentum. Although we
shall not consider in this work an external magnetic field (or vector
potential ${\bf A}$) it can be easily included using the canonical momentum
${\bf P}=-i\hbar\nabla+\frac{e}{c}{\bf A}$ in Eq.\ (\ref{eq1}). 

For GaAs one has $\lambda_R=5.33$ \AA$^2$ and a material-specific
constant $\gamma=27.5$~eV{\AA$^3$} \cite{Kna96}. The latter determines
the Dresselhaus coefficient in terms of the dot vertical width $z_0$ 
as \cite{Vos01} $\lambda_D\approx\gamma(\pi/z_0)^2$ . 
As a representative case, we consider an external parabolic confinement 
with energy $\hbar\omega_0=4.2$~meV and a number of 
electrons ranging from 6 to 12. A strict two-dimensional motion is assumed
and the dot width is simply taken as a parameter for the Dresselhaus 
coupling.
The Rashba term turns out to be negligible in all cases
we have studied. However, the Dresselhaus term increases its importance for 
decreasing doth widths, leading to relevant effects as we shall show below
already for widths $z_0\approx 50$~\AA.
The influence of the Rashba term has already been studied in the literature
for InSb dots \cite{Dar93}, where band non-parabolicity is important, and in 
phenomenologically fitted models \cite{Jac97}.

We have implemented the numerical solution on a spatial grid of the 
full Hamiltonian ${\cal H}_{\em full}={\cal H}_0 + {\cal H}_R + {\cal H}_D$,
with 
\begin{equation}
{\cal H}_0 = \sum_{i=1}^N{\left[\frac{{\bf P}^2}{2m}+
\frac{1}{2}m\omega_0^2 r^2\right]_i}\; .
\end{equation}
However, before presenting numerical results it is worth to 
obtain an approximate analytical solution of the relevant contributions
${\cal H}_0+{\cal H}_D\equiv {\cal H}$ with an approach similar
to that of Aleiner and Fal'ko \cite{Ale01}. A diagonalization to 
second order in $\lambda_D$ is obtained by means of the transformation
$\tilde{\cal H}=U^+{\cal H}U$, where
\begin{equation}
\label{eqU}
U=\exp{
\left[\rule{0cm}{0.5cm}\right.
-i \lambda_D\frac{m}{\hbar^2}\sum_{j=1}^N{(x\sigma_x-y\sigma_y)_j}
\left.\rule{0cm}{0.5cm}\right]
}\; .
\end{equation}
We find
\begin{eqnarray}
\label{eq5}
\tilde{\cal H} &=& \sum_{j=1}^N{ 
\left[\rule{0cm}{0.5cm}\right.
{{\bf P}^2\over 2m} + \frac{1}{2}m\omega_0^2 r^2}
+ \lambda_D^2 \frac{m}{\hbar^3}\, ( x P_y - y P_x )\, \sigma_z 
\left.\rule{0cm}{0.5cm}\right]_j\nonumber\\
&-&N \lambda_D^2\frac{m}{\hbar^2}
+ O(\lambda_D^3)\; .
\end{eqnarray}
Identifying the orbital angular momentum $\ell_z=xP_y-yP_x$ 
in (\ref{eq5}),
we realize that the Dresselhaus term introduces a correction depending 
on $\ell_z\sigma_z$
quite similar to that of the Rashba term for a circular system, 
but in the transformed reference frame. In fact, in the laboratory frame 
the ${\cal H}$ eigenstates to first order in $\lambda_D$ are spinors
that slightly deviate from the $\sigma_z$ eigenspinors. 
We shall call them quasi-up $\chi_{qu}$ and quasi-down $\chi_{qd}$ states
\begin{eqnarray}
\label{eq6}
\chi_{qu}&\equiv&\varphi_{n\ell}({\bf r})
{1 \choose -\lambda_D\frac{m}{\hbar^2} (y+ix)}\; , \nonumber\\
\chi_{qd}&\equiv&\varphi_{n\ell}({\bf r})
{\lambda_D\frac{m}{\hbar^2} (y-ix) \choose 1}\; ,
\end{eqnarray} 
where $\varphi_{n\ell}({\bf r})$ are the eigenfunctions of the circular 
harmonic oscillator ${\cal H}_0$.
Since the dipole operator is invariant by the transformation (\ref{eqU}),
the analytical solution predicts dipole
transitions with $\Delta\ell=\pm1$ 
for quasi-up and quasi-down subsystems (see Fig.\ 1) at the energies
\begin{eqnarray} 
\hbar\omega_{qu,\pm 1}&=&\hbar\omega_0\pm\lambda_D^2\frac{m}{\hbar^2}\nonumber\\ 
\hbar\omega_{qd,\pm 1}&=&\hbar\omega_0\mp\lambda_D^2\frac{m}{\hbar^2}\; . 
\end{eqnarray}
An inmediate consequence is the splitting of the
original Kohn mode at $\hbar\omega_0$
in two peaks separated by an energy 
$\Delta\hbar\omega=2\lambda_D^2 m/\hbar^2$.

Left panels of Fig.\ 2 show the numerical result for the 
FIR absorption of ${\cal H}_{\em full}$ obtained within a real-time simulation 
after an initial rigid displacement of the system as in Ref.\ 
\cite{Pue99}. The results correspond to $N=6$ and 
10 electrons with a Dresselhaus parameter for a 
dot width of $z_0=50$~\AA.
The splitting around $\hbar\omega_0$ nicely agrees with 
the analytical prediction (arrows), thus proving the validity of the model.
For the $N=10$ case we note the existence at low energy of a peak
with much less strength than the dominant ones (notice the enlarged
scale). This too can be explained within the analytical model 
realizing that $N=10$ has a partial occupation of the third
oscillator major shell and thus the possibility of quasi-spin flip
transitions $\chi_{qu}\leftrightarrow\chi_{qd}$ at an energy
$4\lambda_D^2 m/\hbar^2$, indicated with dotted lines in Fig.\ 1
and with an arrow in Fig.\ 2.
Note that in spite of having
$\Delta\ell_z=0$ these transitions can manifest in the dipole spectrum 
through the higher order corrections in Eq.\ (\ref{eq5}).

Due to the spin-orbit coupling the charge dipole oscillation also induces 
a spin oscillation. This is proved in the right panels of Fig.\ 2 where 
the Fourier transform of $\langle S_x +S_y\rangle$ is plotted. The same
frequencies of the dipole absorption manifest in the spin channels, although
with different relative strengths. It is also interesting to analyze the 
spatial patterns of induced density $\delta\rho({\bf r},t)$,  
parallel spin $\delta {\bf S}_\parallel({\bf r},t)$ and vertical spin 
$\delta S_z({\bf r},t)$ for a given oscillation mode and a fixed 
time (Fig.\ 3), where we define the spin density as
${\bf S}({\bf r}) =
\sum_{i=1}^N{\langle \chi_i\, |\, 
{\vec\sigma}\delta({\bf r}-{\bf r}_i)\, |\, \chi_i\rangle }$.
This local-signal analysis has been performed using the method of Ref.\ 
\cite{Val01}. Quite remarkably, the oscillation of $\delta S_z({\bf r},t)$
is always in a perpendicular direction to that of the charge density, 
which is explained in the analytical model as a coherent quasi-up and 
quasi-down excitation due to the SO term.
On the contrary, 
the parallel spin patterns depend on whether the mode has quasi-spin flip 
character or not. For the split Kohn modes (peaks a and b)   
$\delta {\bf S}_\parallel({\bf r},t)$ is localized in the two regions 
perpendicular to the charge oscillation axis. However, 
the quasi-spin flip peak (c) shows no angular dependence of the pattern.
It can be shown that this behaviour is in complete agreement with the prediction 
of the analytical model using the spinors of Eq.\ (\ref{eq6}). 

We consider next how the features discussed above are modified by 
the addition of the Coulomb interaction 
\begin{equation}
\label{eq7}
{\cal H}_C=\sum_{i<j}^N{e^2\over \kappa|{\bf r}_i-{\bf r}_j|}\; ,
\end{equation}
where $\kappa$ is the semiconductor dielectric constant ($\kappa=12.4$
for GaAs). We shall estimate interaction effects using a 
selfconsistent mean-field for the Hartree electrostatic potential
\begin{equation}
V_H({\bf r})= \frac{e^2}{\kappa}
\int{d{\bf r}' {\rho({\bf r}')\over |{\bf r}'-{\bf r}|}}\; ,
\end{equation}
and the LSDA for electronic exchange and correlation. Within the 
present spinor formalism the dots have a vectorial
spin magnetization ${\bf S}({\bf r})$ and, therefore, a general 
exchange-correlation functional of the type
$E_{XC}[\rho,{\bf S}]$ is necessary. For this we have resorted to the 
non-collinear-spin formulation of LSDA theory, described for instance
in Ref.\ \cite{Hei99}, and employed the Monte-Carlo results of 
Ref.\ \cite{Tan89} for the bulk electron gas.

The modifications stemming from the Coulomb interaction are twofold;
first, the electron-hole transitions move in energy due to the change in 
the effective mean field and; second, new collective peaks 
may appear as a result of coherent electron-hole excitations.
Figure 4 shows the interacting results for $N=6$ with 
$z_0=50$ and 35~\AA. Note that the splitting of the Kohn-like 
modes disappears for the larger width and, at the same time, a collective peak
begins to appear at an energy below the electron-hole states.  
These results clearly show the formation of the collective plasmon and magnon states
lying above and below the electron-hole transitions, respectively. 
We stress that in the absence of SO coupling the magnon state 
does not contribute to the FIR absorption.  
The lower left panel of Fig.\ 4 indicates that a great enhancement of 
the magnon contribution
is obtained by reducing the dot width, which is accompanied by an
important Landau damping of the plasmon state into a bundle of peaks.   
The Coulomb interaction thus suppresses 
the splitting of the Kohn-like modes although it introduces the mechanism 
of Landau damping that ultimately, and rather abruptly with $z_0$, 
dominates for decreasing widths.

The right panels of Fig.\ 4 show the horizontal spin spectra induced by the 
dipole shift. These 
results permit to ascertain the spin character of the different modes and, by 
comparison with the left panels, quantify their relevance to the FIR
absorption. The relative spin or density character of each state
is also seen from the amplitudes ${\cal A}$ of density and spin oscillation. 
For instance,  when $z_0=50$~{\AA} we find 
${\cal A}[\delta S_z({\bf r},t)] \approx 60\, 
{\cal A}[\delta\rho({\bf r},t)]$ for the 
magnon and 
${\cal A}[\delta S_z({\bf r},t)] \approx 0.4\, 
{\cal A}[\delta\rho({\bf r},t)]$ for the plasmon, 
which shows the 
dominant spin and density type for magnon and plasmon states, respectively.

The characterization of the interacting peaks is facilitated
by the comparison of the spatial patterns with the corresponding ones 
in the non-interacting model. Figure 5 displays the density and spin 
patterns for two of the peaks of the $z_0=35$~{\AA} spectrum
(lower left panel of Fig.\ 4).
As in the non-interacting model, the oscillation of $\delta S_z({\bf r},t)$ 
is orthogonal to that of $\delta\rho({\bf r},t)$ and the horizontal spin 
patterns show a priviledged spin direction, i.e., they have a net 
induced spin. Note in particular that peak b of Fig.\ 4 has no 
angular dependence in the horizontal spin pattern while for peak
a there are two regions with enhanced $\delta {\bf S}_\parallel({\bf r},t)$.
We have checked that these features are not changed
when the polarization of the mode, given by the 
direction of the density oscillation, is varied.
We stress that the mechanism discussed here would permit a control of 
the horizontal spin oscillation by tuning the frequency of the 
applied electric field. This effect of spin-to-charge conversion 
is a major requisite for hybrid spintronic devices \cite{Wol01}.

Finally, Fig.\ 6 presents the spectra with Coulomb interaction 
obtained for $N=10$, an open-shell system in the
analytical model. 
At $z_0=50$~{\AA} the open-shell character of 
this dot is clearly maintained, with low-energy electron 
hole transitions and an associated collective mode
at $\omega\approx 0.7$~meV similar 
to the quasi-spin flip excitations of Fig.\ 1. For narrower dots, however,
the energy of quasi-spin flip transitions increases and as a result this mode 
merges with the magnon and other electron-hole states. 
Contrary to the quasi-spin flip state, both magnon and plasmon 
mean energies are not dependent on the Dresselhaus coupling. 

To summarize, the role of the SO coupling in the FIR absorption
of GaAs dots has been studied. 
In an analytical model without interaction the relevant Dresselhaus 
contribution induces characteristic splittings and quasi-spin flip modes. 
The Coulomb interaction favors the formation of collective plasmon and 
magnon states and it suppresses the 
splittings of the analytical model. Nevertheless, below a certain dot 
width Landau damping dominates the spectrum.
Each mode has characteristic patterns of induced charge, horizontal spin 
and vertical spin, as well as a ratio of oscillating density
and spin amplitudes.
Therefore, by tuning the frequency of the applied electric
field one can effectively control the dynamical spin properties
and the relative intensities of the charge and spin oscillation. 

This work was supported by Grant No.\ PB98-0124 from DGESeIC, Spain.

\begin{figure}[b]
\centerline{\psfig{figure=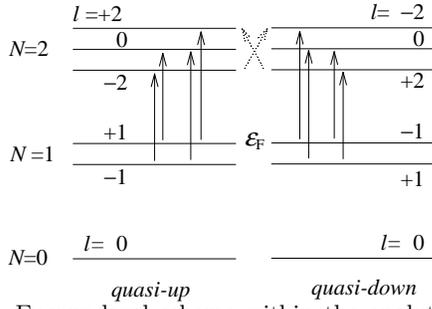,width=2.25in,clip=}}
\caption{Energy level scheme within the analytical model
for ${\cal H}_0+{\cal H}_D$. 
The Fermi
level for $N=6$ is indicated by $\varepsilon_F$ and the vertical 
arrows correspond to the 
allowed transitions having $\Delta\ell=\pm1$ within each quasi-spin 
subset.
The dotted lines show quasi-spin flip $\Delta\ell=0$ transitions 
within the $N=2$ major shell, discussed in the text for $N=10$.}
\end{figure}

\begin{figure}[h]
\centerline{\psfig{figure=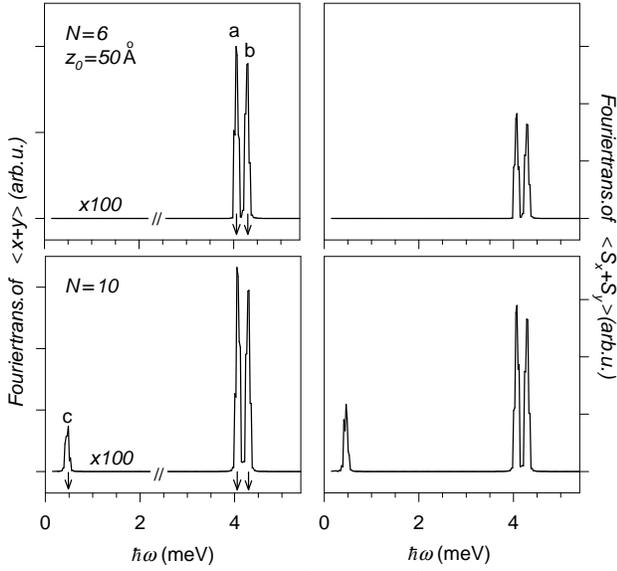,width=3.2in,clip=}}
\caption{
FIR absorption (left panels) obtained from the Fourier
transform of the real time dipole oscillation 
with ${\cal H}_{\em full}$ for a dot width of $z_0=50$~\AA.
Right panels display the corresponding 
frequency transform of the horizontal spin signal. The arrows
indicate the energies of the split Kohn modes
(at $\approx 4$~meV) and of the quasi-spin flip modes  
(at $\approx 0.5$~meV) using the analytical model for this $z_0$.
}
\end{figure}

\begin{figure}[b]
\caption{
Spatial patterns of oscillating density $\delta\rho({\bf r},t)$,
horizontal spin $\delta{\bf S}_\parallel({\bf r},t)$ and vertical spin
$\delta S_z({\bf r},t)$ at a fixed time for the a and c dipole modes
of Fig.\ 1. The gray color scale of left and right panels indicates
the signal magnitude while the corresponding sign is superimposed
in white.   
}
\end{figure}

\begin{figure}[h]
\centerline{\psfig{figure=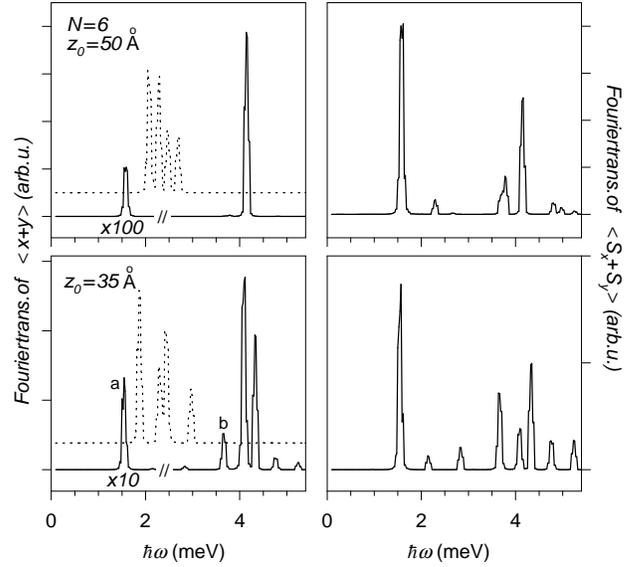,width=3.2in,clip=}}
\caption{Same as Fig.\ 2 including Coulomb interactions. The dashed line
in the left panels displays the electron-hole transitions in the 
selfconsistent mean-field.}
\end{figure}

\begin{figure}[h]
\caption{
Same as Fig.\ 3 including Coulomb interaction. Upper and lower row
correspond to the peaks labeled a and b in Fig.\ 4, respectively.}
\end{figure}

\begin{figure}[h]
\centerline{\psfig{figure=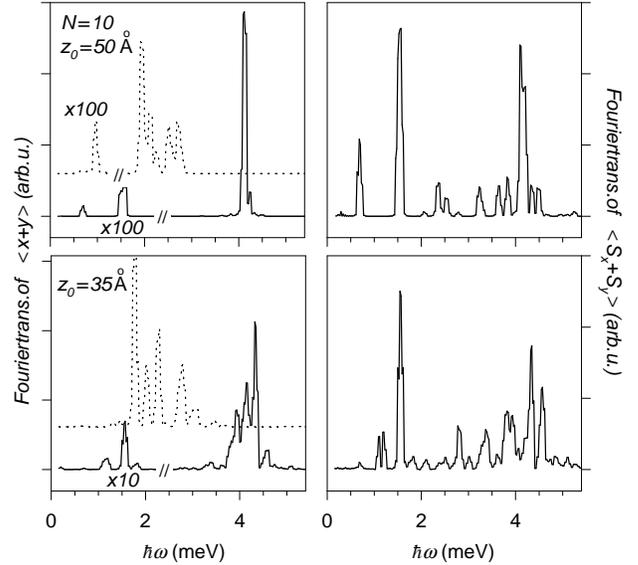,width=3.2in,clip=}}
\caption{
Same as Fig.\ 4 for $N=10$.}
\end{figure}

\end{document}